# ENERGY RAMPING FOR DAFNE-2

C. Milardi, LNF-INFN, Frascati, Italy

*Abstract*

The aim of this paper is to overview the energy ramping main issues for DAFNE-2, the high-energy upgrade of the Frascati $e^+e^-$ DAΦNE collider in order to understand the feasibility of the process itself and to define the tools to be developed for its implementation.

## INTRODUCTION

Two possible evolutions have been considered for the DAΦNE [1] future. The first scenario assumes a completely new machine providing collisions at the energy of the φ-factory, 1.02 GeV in the center of mass, with a luminosity at least two order of magnitude higher than in the present DAΦNE configuration, which has reached a peak luminosity of $.8 \ 10^{32}$ cm$^{-2}$ s$^{-1}$. This machine should be of interest for high-energy physics studies especially in the field of kaon rare decays and the related flavor physics.

A second scenario considers a smooth upgrade of the existing collider to increase the collision energy in the range of 2 ÷ 2.4 GeV in the center of mass; it offers the opportunity to investigate events close to the neutron-antineutron threshold and improve the statistics about the nucleons form factor measurements. The so called DAFNE-2 [2] option is much less challenging and much less demanding in terms of design, manpower, realization time and cost, especially if the modifications are limited to the bending magnets and to the interaction region and if the existing injection system![3], see Fig. 1, can be reused injecting the beam at .51 GeV energy and then ramping the rings to the colliding energy.

## DAFNE-2 GENERALITIES

DAFNE-2 uses only one out of the two available interaction regions.

Each ring is supposed to work in the energy range 1 ÷ 1.2 GeV, circumstance asking for an upgrade of the bending magnets [4]; in fact the existing dipoles can provide a maximum magnetic field equal to ≈ 1.8 T while 2.35 ÷ 2.8 T will be required unless the present bending radius (1.4 m) is not increased. Such magnetic field variation corresponds to a ≈ 520 A current increase on the power supplies, assuming to reuse the same devices presently working on DAΦNE.

The design of the low-β section relies on two super-conducting quadrupole doublets; each quadrupole is installed with a proper rotation angle in order to follow the rotation of the betatron oscillation plane introduced by the detector solenoid having a field integral of 0.7 Tm.

A luminosity in the range of $10^{32}$ cm$^{-2}$ s$^{-1}$ has been required for the experiments proposed for the high energy update of DAΦNE [5]. This luminosity, relying on DAΦNE experience, can be obtained with currents of the order of .5 A per beam stored in 30 buckets, out of the 120 available, with a rather low current per bunch. In this context the rings can be even operated without wiggler magnets.

Wigglers had been introduced in the DAΦNE lattice to increase damping, which helps in fighting beam-beam driven instabilities as well as multi-bunch effects. Since the wigglers operate at their maximum field already at 510 MeV, a lattice without wigglers can be ramped at a constant optical configuration, limiting the ramping process to a simple synchronous scaling involving all magnetic elements. In presence of wiggler magnets an efficient approach would require interleaving each ramping step with slow feedbacks for closed orbit and betatron tunes.

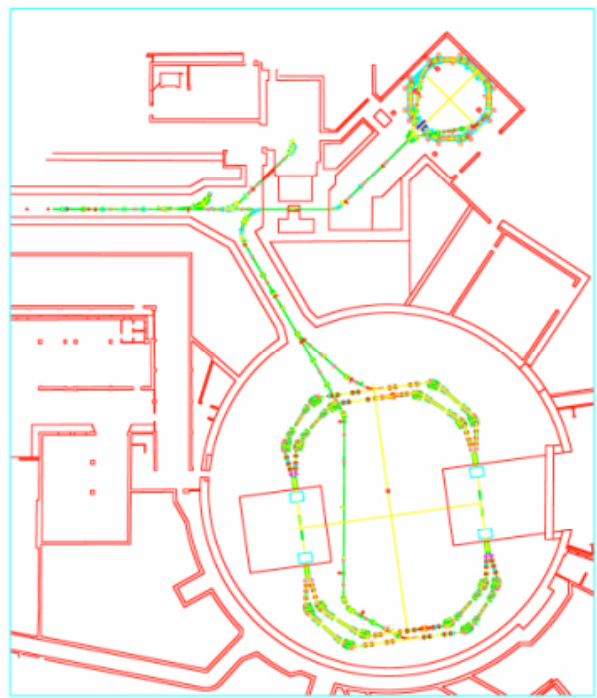

Figure 1: Present DAΦNE layout showing the $e^+e^-$ colliding rings and the injection system.

## INJECTION STRATEGY

In principle the energy ramping procedure in a collider such as DAFNE-2 can be implemented in three simple steps:

- separation of the two beams at the crossing point
- synchronous ramping of both rings from injection energy to the desired interaction one
- removal of beam separation.

Beam separation can be realized in the longitudinal plane by a fast phase jump of the RF cavity in one ring.

This method [6] has been tested and used routinely during DAFNE operation before finding an efficient procedure for injection in collision. It has proved to be reliable with trains of 30 and 40 bunches and jumps of 1 and 1.5 buckets.

## RAMPING MAIN ISSUES

The main issue in an energy ramping process is to change the accelerator energy in a time as short as possible preserving the beam current and stability while keeping tunes, closed orbit and chromaticity constant.

A stable beam current during ramping is necessary to maintain the best collider performances in term of peak and integrated luminosity, limiting the background hitting the detector. Longitudinal and transverse beam stability must be preserved as well to avoid beam loss and the related background showers.

Last but not least ramping at fixed betatron tunes minimizes the time necessary to set up the rings for collision.

### Elements involved in the ramping process

In a collider the ramping process involves all magnetic elements (with the related physical quantities); for DAΦNE they are:

- Dipoles (reference orbit $x_{ref}$ and $y_{ref}$.)
- Quadrupoles (betatron tunes $\nu_x$ and $\nu_y$)
- Sextupoles (chromaticity $\xi_x$ and $\xi_y$)
- Splitters (horizontal crossing angle at the interaction point $\theta_{cross}$)
- Steering magnets (closed orbit)

At DAΦNE four different kinds of bending magnets are installed: sector and parallel short [7] and sector and parallel long [8] providing respectively 40.5° and 49.5° deflection angles while ordinary quadrupoles are of three different kinds, small [9], large [10] and large aperture [11].

In practice the magnetic field, for each ramp step $i$ and for each element $j$ must be changed by setting the current provided by the corresponding power supply according to the calibration curve specific to that element:

$$B_{ji} = f(I^{PS}_{ji}) \qquad i=1,\ldots n_{step} \qquad j=1,\ldots n_{ele}$$

Calibration curves for long and short dipoles and for small quadrupoles are presented in Figs. 2, 3 and 4 respectively.

Dedicated High Level Software (HLS) applications, running within the collider Control System, will be necessary to compute $I^{PS}$ arrays to implement the ramping process.

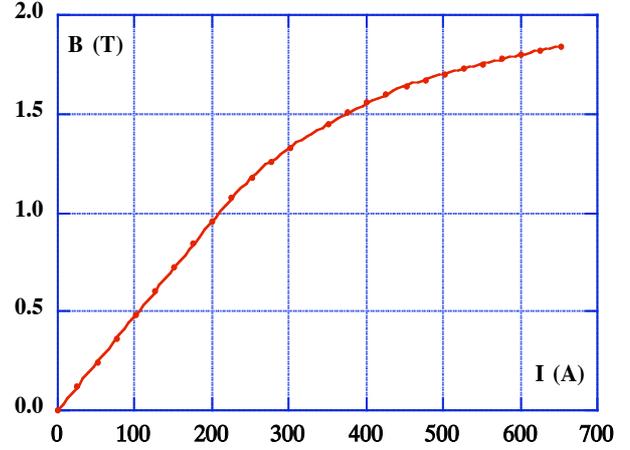

Figure 2: Vertical component of the field, at the magnet center, as a function of the coil current for the DAΦNE long dipoles.

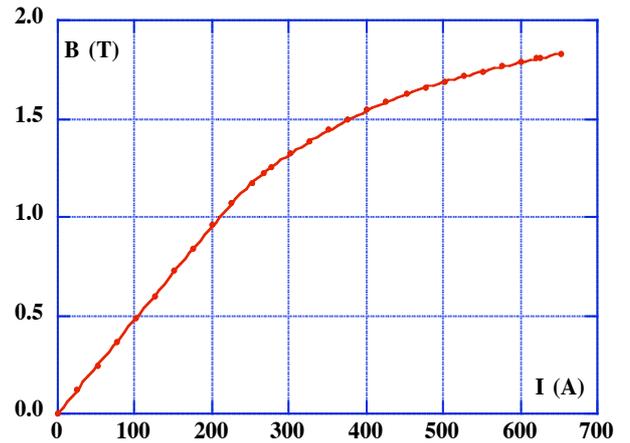

Figure 3: Vertical component of the field at the magnet center as a function of the coil current for the DAΦNE short dipoles.

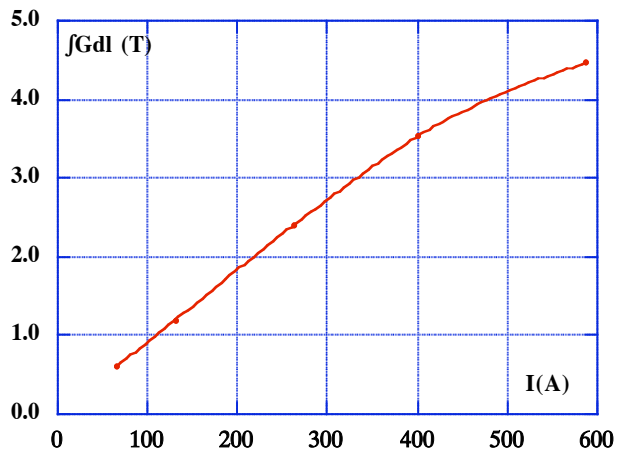

Figure 4: Average integrated gradient versus excitation current for the DAΦNE small quadrupoles.

*Wigglers and ramping*

Beam stability at injection energy could be improved by keeping the wigglers on. For this reason we still keep this option for DAFNE-2, although it would have a strong incidence on the power bill and would add some complications to the ramping process. In this case the wiggler magnets will be operated at fixed field, while ramping, and the tune and orbit shift during ramping will be recovered by implementing slow feedbacks for the closed orbits and the betatron tunes to be included in the ramping procedure.

*Coupling and ramping*

The ramping option requires a design for the DAFNE-2 IR based on superconducting quadrupoles instead of the present system of permanent magnet ones. The low-beta quadrupoles are rotated by the corresponding average rotation introduced by the detector solenoid at their azimuth to minimize the betatron coupling. Experience with DAΦNE has shown that a careful coupling correction is mandatory to reach the best performances in terms of luminosity [12]. The design value for DAFNE-2 $\kappa = 0.3$ % has already been reached at DAΦNE.

Special attention deserves coupling evolution: in fact it will be no longer compensated neither at the injection nor during the ramping process due to the practical difficulty in rotating continuously the superconducting low-β quadrupoles. This point does not seem relevant to beam stability because the two beams are ramped out of collision and at constant betatron tunes to avoid resonances. In terms of beam lifetime a larger coupling during ramping could even introduce some advantages since lifetime is Touschek dominated and, as a consequence, strongly dependent on the bunch size.

It is worth recalling that the Touschek lifetime depends on the inverse cubic power of the beam energy and for this reason it is much shorter at injection energy.

Therefore, a longer lifetime during ramping could be helpful in improving the integrated luminosity. However, betatron coupling correction can be assured, even during ramping, adding skew quadrupole windings to the permanent magnet quadrupoles, or implementing coupling correction procedure based on skew quadrupoles.

*Ramping speed and Synchronization*

The average ramping speed for a given magnetic element is defined as:

$$R_{speed} = \frac{I_{E\,coll} - I_{E\,inj}}{\Delta t}$$

$R_{speed}$ must be evaluated for each class of elements taking into account the calibration curve so to implement the same energy variation in the same time range, and having care, at the same time, to avoid eddy current related problems.

All DAΦNE elements are laminated and therefore the main limitation to the ramping speed comes from the vacuum chamber.

The slew rate of the dipole power supply can be tuned in the range:

$$sr = 7.5 \div 75 \text{ A/s}$$

corresponding, for a ≈ 520 A current variation, to a ramping time $t_{ramp}$

$$t_{ramp} = 7. \div 70 \text{ s}$$

The maximum ramping speed is not compatible with the characteristics of the DAΦNE vacuum chamber, while a slower ramp, up to 40 s, would not create any problem.

However, in order to linearize the different calibration curves of the magnetic elements between injection and operation energy it will be necessary to implement ramping process in steps.

*Ramping time evaluation*

In order to estimate the time necessary to inject from scratch 0.5 A in each beam and reach the colliding condition, we must take into account that:

> 30 steps are a quite reasonable choice to linearize the different calibration curves and a $t_{ramp} \approx 2$ s is a conservative assumption for varying the dipole power supply current.
> 
> at each step 2 s are required for the slow orbit and betatron tunes feedbacks
> 
> 30 s are needed to inject 0.5 A in each beam
> 
> the injection system requires 180 s to switch from electron to positron mode

the typical operation will be:

> ramp down from operation to injection energy: 120 s (2 s for the magnets + 2 s for the slow feedback a step)
> 
> electron injection: 30 s
> 
> switch from electrons to positrons: 180 s
> 
> positron injection: 30 s
> 
> ramp up to operation energy: 120 s

The total time is ≈ 8 minutes. This time is quite reasonable if compared with the estimated beam lifetime in collision, of the order of 2 h [13].

*Power supply ramp synchronization*

Although the DAΦNE power supplies are not used presently in a synchronous configuration, synchronized ramps can be implemented within the accelerator Control System [14] since all the power supplies have a remotely settable slew-rate and accept an external hardware trigger. So, once the proper slew-rates have been computed by the HLS procedure and applied to the corresponding power supply by a command executed by a remote CPU, the synchronized ramp can be started by sending a trigger signal, see Fig. 5.

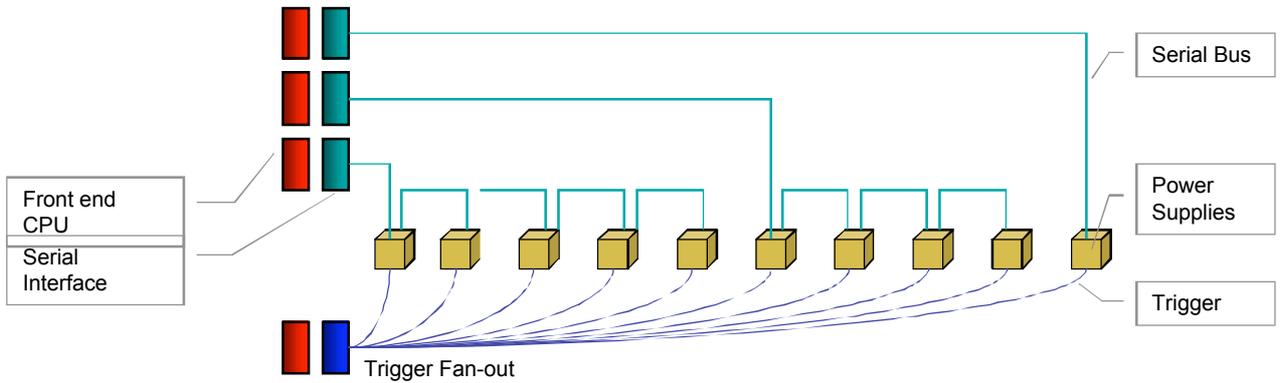

Figure 5: Power Supply ramping synchronization scheme.

## TRANSVERSE AND LONGITUDINAL FEEDBACKS

The DAΦNE colliding rings are equipped with transverse [15] and longitudinal [16] feedbacks to cope with the effects of coupled-bunch instabilities.

The energy ramping process does not affect transverse feedback operation as long as the beams are ramped without changing the orbit position, in particular at the pick-up where the bunch-by-bunch offset is detected. Moreover they will be also useful to damp possible coherent oscillations driven by slightly negative chromaticity.

Special attention deserves the longitudinal feedback, which is sensitive to the variation of the synchrotron oscillation frequency $\nu_s$ induced by the energy ramping. The simplest solution for this problem consists in compensating the $\nu_s$ shift by adjusting the Radio Frequency voltage.

However it might be even possible to exploit the adaptive feature provided by the feedback system itself, able to scan eight different kinds of filters in real time. Yet more sophisticated systems providing advanced adaptive features, presently under study [17], can be considered if necessary.

## SLOW FEEDBACKS

### Orbit

The DAΦNE closed orbit acquisition system samples the beam position at a rate of 6 Hz [18]. Without further improvements it will be possible to implement a slow orbit feedback working at the rate of 2 Hz.

The beam position monitor readouts are acquired by four parallel remote processors, which elaborate the information from the BPMs, giving the horizontal and vertical beam position. A CPU at the second level of the Control System reconstructs the beam orbit from these raw data taking into account the BPM alignment offsets, the reference system transformation affecting BPMs installed within the detector solenoid and provides averaging; the result is displayed on the user interface.

Presently the closed orbit correction works under the operator control, and the measured response matrix $A_m$ of the steering magnet deflections $\overline{\theta}$ is used to find, by means of the singular value decomposition numerical technique, the beam orbit $\overline{z}$ [19].

$$\overline{z} = A_m \Delta\overline{\theta}$$

This orbit correction method can be easily included in an automated process running at the second level of the Control System under the control of the ramping procedure.

### Betatron tunes

The DAΦNE betatron tune measurement system is being upgraded exploiting the hardware developed for the beam transverse feedback. It delivers a fast analog to digital sampling of the signal from a four buttons beam position monitor [20]. Once operative this system will provide, by injecting a bunch out of collision, a continuous real-time measurement of the tunes for the two beams. Moreover it will be a natural front end for a betatron tune slow feedback consisting in an application able to measure the tunes, to compute the proper quadrupole configuration compensating the tune shifts and to restore the initial tune values by varying the quadrupole power supplies set points. The quadrupole configuration will be computed at each step by means of the machine model integrated in the Control System and routinely used for optics computation, fine tuning and energy scan [21].

## CONCLUSIONS

The energy ramping process for DAFNE-2 does not show any special problem in principle. It asks only for a careful hardware configuration and an intense HLS applications development in order to implement the required ramping procedures and to develop the slow feedbacks on closed orbit and betatron tunes.

The time necessary to execute the whole ramping process has been estimated to be 8 minutes, which is quite negligible if compared with the expected beam lifetime in collision.

For a final evaluation of the total ramping time it is necessary to frozen the design of the new dipoles and that of the superconducting quadrupoles in the IR.

## AKNOWLEDGMENTS

The author wishes to thank C. Biscari, A. Drago, M. Preger, R. Ricci, M. Serio, A. Stecchi for many useful discussions.